# GPU ACCELERATED AUTOMATED FEATURE EXTRACTION FROM SATELLITE IMAGES

K. Phani Tejaswi, D. Shanmukha Rao, Thara Nair, A. V. V. Prasad

National Remote Sensing Centre, Indian Space Research Organization, Hyderabad, India
tejaswi_phani@nrsc.gov.in, thara_nair@nrsc.gov.in

## ABSTRACT

*The availability of large volumes of remote sensing data insists on higher degree of automation in feature extraction, making it a need of the hour. Fusing data from multiple sources, such as panchromatic, hyper spectral and LiDAR sensors, enhances the probability of identifying and extracting features such as buildings, vegetation or bodies of water by using a combination of spectral and elevation characteristics. Utilizing the aforementioned features in remote sensing is impracticable in the absence of automation. While efforts are underway to reduce human intervention in data processing, this attempt alone may not suffice. The huge quantum of data that needs to be processed entails accelerated processing to be enabled. GPUs, which were originally designed to provide efficient visualization, are being massively employed for computation intensive parallel processing environments. Image processing in general and hence automated feature extraction, is highly computation intensive, where performance improvements have a direct impact on societal needs. In this context, an algorithm has been formulated for automated feature extraction from a panchromatic or multispectral image based on image processing techniques. Two Laplacian of Guassian (LoG) masks were applied on the image individually followed by detection of zero crossing points and extracting the pixels based on their standard deviation with the surrounding pixels. The two extracted images with different LoG masks were combined together which resulted in an image with the extracted features and edges. Finally the user is at liberty to apply the image smoothing step depending on the noise content in the extracted image. The image is passed through a hybrid median filter to remove the salt and pepper noise from the image. This paper discusses the aforesaid algorithm for automated feature extraction, necessity of deployment of GPUs for the same; system-level challenges and quantifies the benefits of integrating GPUs in such environment. The results demonstrate that substantial enhancement in performance margin can be achieved with the best utilization of GPU resources and an efficient parallelization strategy. Performance results in comparison with the conventional computing scenario have provided a speedup of 20x, on realization of this parallelizing strategy.*

## 1. INTRODUCTION

The massive increase in the dimension of remotely sensed data that is being obtained by an ever-growing number of earth observation sensors, has led to a condition, where there is large number of images to be processed. To tide over this scenario, automated feature extraction techniques can be employed for the extraction of geo-spatial features. Automated feature extraction can be defined as the identification of geographic features and their outlines in remote-sensing imagery through post-processing technology that improves feature definition, either by increasing feature-to-

      

background contrast or by the usage of pattern recognition software [1]. This is an important task in numerous applications ranging from security systems to natural resource inventory based on remote-sensing. Image interpretation is an important and highly challenging problem with numerous practical applications.

Edge detection is one of the most commonly used operations in image analysis and there are many algorithms for enhancing and detecting edges .Edges form the boundary between an object and the background, and indicates the boundary between overlapping objects. Traditional approaches use classical edge detectors viz. Sobel operator, Canny Edge detector, Prewitt edge detector and LoG operators that work fine with high quality pictures. Of these the LoG operator is the most commonly used. Different features of interest can be extracted by varying the LoG filter size. Filter of high variance yields the finer edges of the image where as filter of low variance gives the inner details of the image.

Zero crossings lie on closed contours. Zero crossing detectors provide a binary image with single pixel thickness lines indicating zero crossing points. 'Edges' in the images appear as zero crossings and can be detected by sign change in LoG in at least one direction. When the contrast with respect to the neighbors of the pixels is very high, a zero crossing will be detected and can be extracted to identify and demarcate the edges more clearly. The standard median filter [2] is a simple rank selection filter that removes impulse noise. This is achieved by replacing the luminance value of the center pixel of the filtering window with the median of the luminance values of the pixels contained within the window. The drawback of median filters is that, it removes thin lines and blurs image details at low noise densities. Median filtering technique is generally employed to eliminate the salt and pepper noise from an image. The hybrid median filter is a modified median filter that draws the luminance values from the pixels parallel to and at 45deg to central pixel of the filtering window. This has been proposed to avoid the inherent drawbacks of the standard median filter by controlling the tradeoff between the noise suppression and detail preservation [3].

This paper discusses the edge detection, focusing on LoG, zero crossing detection and various statistical techniques to demarcate the edges and to extract the different features in panchromatic data, applicable for urban area detection. Combining these image processing approaches in a panchromatic image or a single band in a multispectral environment enables us to extract features of interest viz. water bodies, roads, buildings etc. The related work carried out earlier in this field is discussed in second section and the theoretical reasoning behind the selection of LoG operator, zero crossing threshold, removal of salt and pepper noise and their applicability in this development are explained in the succeeding section. Fourth section describes the automated feature extraction algorithm and its implementation details. Subsequent sections give an insight on the General Purpose Graphical Processing Units (GPGPUs) and their effective exploitation by adopting various design approaches. Detailed evaluation of the performance of the finalized optimal approach is discussed with the performance margin and the results on various test images.

## 2. RELATED WORK

Feature extraction of satellite images is of major interest in a slew of applications as it is the major input for the remote sensing applications. Edge detection employing techniques is also dominant in various image feature extraction techniques in which 2-D wavelet transforms are applied on the image for several levels [4]. Several existing algorithms on edge detection like Marr-Hildreth edge detector, Canny edge detector [5], make use of Laplacian of Gaussian operator with a single exclusive sigma value. Canny edge detector has been efficiently used in the extraction of features from the satellite images [6]. It allows a pre-processing step of Gaussian smoothing prior to actual application of LoG filter. Median filtering [7] is the most commonly used approach for the removal of salt and pepper noise [8] from the image. The ranking of the neighboring pixels depends on the brightness and the median value is now made the new value of the central pixel in a hybrid median filter. Most of the edge detection techniques and feature extraction algorithms were designed for IKONOS images [9,10] and similar with a very high resolution of the order of 1 m.

The approach discussed in this paper targets Cartosat-1 and Resourcesat AWiFs images which are lower resolution images. The same algorithm can be applied for extracting the features of the satellite images of very high resolutions. The LoG operator followed by zero crossing detection has been used as the edge detection technique where no significant details of the image are lost compared to the wavelet approach. The drawback in the application of wavelet technique is that the output resolution decreases with increase in wavelet levels.

Another unique feature of this algorithm is that it employs the LoG operator with two different sigma values, a high sigma to extract the features of interest and that of low sigma to extract the edges in the image.

Yet another highlight is the selection of hybrid median filter for noise removal. The disadvantage of the conventional median filtering techniques is they tend to erase lines narrower than half the width of the neighborhood. Hybrid median filtering is chosen as the technique for the removal of salt and pepper noise in the image as it gets along the limitations of conventional median filtering technique and involves a three-step ranking process that uses two subgroups.

This is totally unsupervised and can be applied with ease to any satellite image of any resolution unlike many algorithms which employ supervised classification in which the pixels need to be trained a priori.

## 3. SCIENTIFIC RATIONALE

### 3.1. LoG

The Laplacian of an image highlights regions of rapid intensity change and is therefore often used for edge detection. These kernels apply a second derivative measurement on the image and hence are highly susceptible to noise frequencies. Using any standard kernel, the Laplacian can be calculated using standard convolution methods. To counter this, the image is often Gaussian smoothed before applying the Laplacian filter. Smoothing by Gaussian helps to reduce the high frequency noise components ahead of differentiation. The LoG operator detects and extracts the edges from an image. The suitable mask can be selected depending on the requirement to extract the edges or the features of interest. The 2-D LoG function centered on zero and with Gaussian standard deviation σ has the form:

$$LoG(x,y) = -\frac{1}{\pi\sigma^4}\left[1 - \frac{x^2 + y^2}{2\sigma^2}\right]e^{-\frac{x^2+y^2}{2\sigma^2}} \qquad (1)$$

By itself, the effect of the filter is to highlight edges in an image.    also controls the size of the Gaussian filter. A higher value for    results in a larger size of Gaussian filters. This implies detecting larger edges. Smaller values of    imply a smaller Gaussian filter maintains finer edges in the image. The selection of variance and window size can be used to provide edges at various scales. LoG operator omits low and high pass frequencies, it is equivalent to band-pass filter.

Choice of the value of variance controls the spread of operator in the spatial domain, setting variance to a high value gives low-pass filtering, as expected.

The advantage of employing LoG operator in edge detection is that it helps in finding the correct places of edges and testing wider area around the pixel. The user can tailor the algorithm by adjusting the variance to adapt to different environments.

### 3.2. Zero crossing

An approach to finding edges is to detect zero-crossings in second derivative or LoG filtered output. The zero crossing is computed by observing the four neighbors of each pixel. If they all have the same sign as candidate pixel, then no zero crossing is detected. On the contrary, candidates having the smallest absolute value compared to its neighbors with opposite sign are indicative of zero crossings. Subsequently the zero-crossings are to be passed through a threshold to retain only those with large difference between the positive maximum and the negative minimum, so as to suppress the weak zero-crossings most likely caused by noise.

### 3.3. Standard Deviation

Sample standard deviation is defined by the unbiased estimate of the standard deviation, $s_a$, of the brightness within a region ($\Re$) with pixels and is denoted by:

Standard deviation of a pixel in an image in general can be computed as:

$$s_a = \sqrt{\frac{\Sigma(a[m,n] - m_a)^2}{\Lambda - 1}} \qquad (2)$$

where $m_a$ is the mean of all values in the data set.

The standard deviation plays an important role in identifying the edges of the image which has been threshold with a zero crossing detector.

### 3.4. Median Filtering

Median filters are non-linear digital filters known for their capability to remove impulse noise, with minimum signal distortion and without hampering the edges. In particular, compared to the smoothing filters examined thus far, median filters have many advantages viz. no reduction in contrast across steps, no shifting of boundaries and no occurrence of new unrealistic values as new values are not created near edges. The degradation of edges is minimal in median filtering, which makes it possible to apply median filters repeatedly. A specialized median filter is the Hybrid Median filer. This adopts a 3-step ranking process using two subgroups of a 5x5 neighborhood. These subgroups are extracted from the pixels parallel and at 45° to the edges, with reference to the center reference pixel. The median for each subgroup is determined. Comparison is carried out between the two values and the original pixel value and the median of these three values forms the output value for the pixel in the selected window in the filtered image. The window size can be selected as required by an application.

## 4. ALGORITHM FOR AUTOMATED FEATURE EXTRACTION

An algorithm which employs image processing techniques has been formulated which enables us to extract information like roads, buildings or water bodies from any panchromatic image or any single band of a multispectral image.

The Laplacian of Gaussian filter is used to extract the edges. Depending on the value of variance, the size of the Gaussian filter varies. A larger Gaussian filter detects larger edges and a smaller variance can be employed to detect finer edges in the image. Hence incorporating a proper combination of these two filters helps us to detect all edges in an image.

Applying a LoG filter alone will not provide edges with the required clarity. Once the edges are filtered through the LoG filter, a zero crossing threshold needs to be applied on the LoG filtered image. The zero crossing detector with a predefined threshold scans through the four neighbors of each pixel. Pixels with the smallest absolute value compared to its neighbors with opposite sign are qualified as zero crossings, once they are cleared through a threshold detector. This stage defines the larger and finer edges of the image, which is to be followed by identifying and marking the required features explicitly.

Identifying and segregating the region of interest was the succeeding requirement. Since the zero crossing threshold detectors was applied on the output of LoG filter, statistical techniques like standard deviation had to be applied on matrices with specified dimensions to isolate the larger and finer edges. The concept of variation in standard deviation values was used to isolate the presence or absence of edges in a 5x5 pixel neighborhood.

The statistical output might have a large presence of salt and pepper noise, which has to be filtered properly. Median filter is the best non-linear digital filter since they remove impulse noise without damaging the edges. Repeated invoking of hybrid median filter with varying dimension and a three stage ranking process yields an image with trivial noise content.

## 4.1. Implementation Details of Urban Area Detection Algorithm

The input image was padded with 2 rows/columns on all the boundaries so that each pixel in the original image had its corresponding 5x5 neighborhood. The Laplacian of Gaussian filters of variance 0.5 and 20 were applied on the image separately to extract the features and edges respectively. The LoG filter dimension selected in both the cases was 5x5. The LoG mask was applied on each pixel with its 5x5 neighborhood and the LoG output images were obtained for the two filters. A zero crossing detector with a preset threshold value was applied on both the LoG output images which were padded with 1 row/column on all the boundaries. The zero crossing logic was applied on each pixel and its corresponding up, down, left and right neighbors. On the application of the zero crossing detectors, features of interest were demarcated in the low variance filtered image and finer edge transitions were identified in the high variance filtered image. Standard deviation value was worked out for each pixel in its 5x5 neighborhood on the zero crossing output images, padded with 2 rows/columns on the boundaries. If the standard deviation exceeded the corresponding pre-specified maximum threshold deviations, the standard deviation was re-calculated for the corresponding pixel with a localized 3x3 neighborhood. Pixels which cross the set threshold deviation were identified. These pixels denote the features of interest in the high variance filtered image where as they symbolize the edges in the low variance filtered image. The features and edges extracted from the low and high variance filtered images respectively were combined together to achieve the urban area mapped image. The steps involved in urban area detection are given in the flow chart below

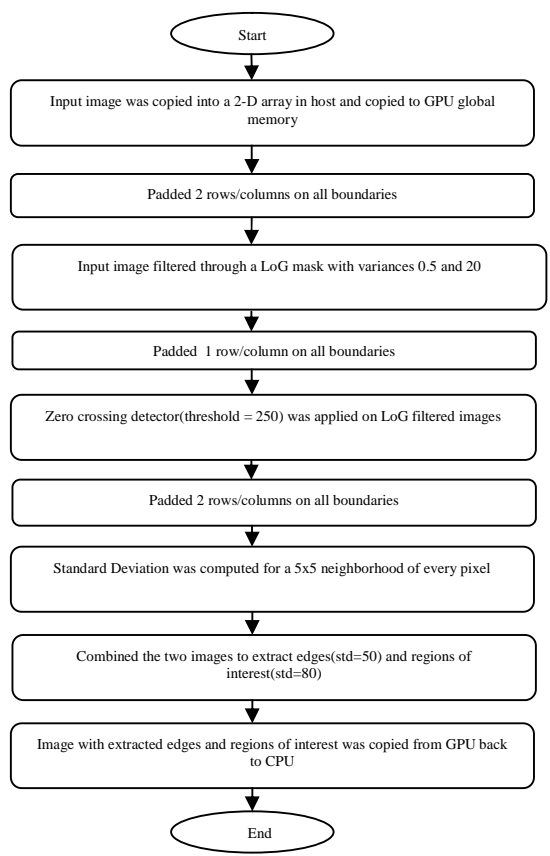

Figure 1. Flow Chart for Urban Area Detection

## 4.2. Implementation Details of Water body ExtractionAlgorithm

The urban area detection algorithm was tailored for the extraction of water bodies in any band of a multispectral image. The feature extracted image has a characteristic salt and pepper noise which was removed by passing through a hybrid median filter in multiple levels of higher and lower dimensions. The steps involved in water body extraction are given in the flow chart below

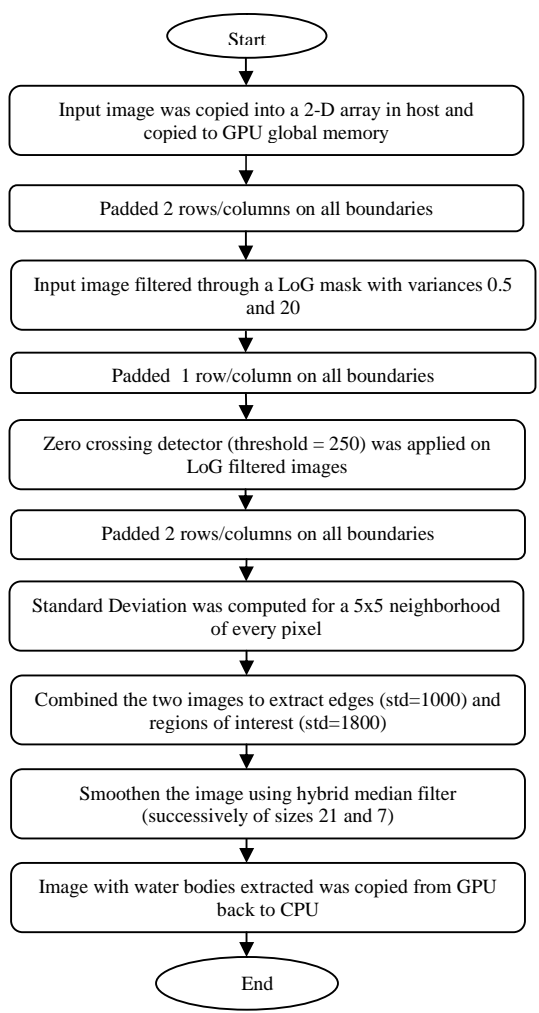

Figure 2. Flow Chart for Urban Area Detection

## 5. GPU- INTRODUCTION, NEED AND OPTIMIZATION STRATEGIES

It is cumbersome to manage the large volume of remote sensing data to be processed using automated feature extraction. For larger images, where significant performance improvement may be required, computation making use of traditional processors may not yield the desired performance. As a result, deployment of GPUs (Graphics Processing Units) has attained momentum in applications where large volume of data needs to be processed. In this context, the application of General Purpose Computing using Graphical Processing Unit (GPGPU), a coprocessor to CPU, for automated feature extraction is being thought of. With remarkable improvements in GPU computing, due to the improved hardware potential and programmability, this paper depicts how a GPU, with parallel design architecture can enhanced computational power significantly to match the requirements of handling enormous data volume[11].

The programming platform that has been selected is NVIDIA Corporation's CUDA (The Compute Unified Device Architecture) [12]. The most important advantage of CUDA over most other GPU-APIs is that it can easily access the memory of GPUs and also permits direct access of the global and shared memory.

Well-organized implementation of an algorithm in the GPU involves extracting the parallelism of the target application, and, additionally, adopting efficient data transfer, memory and thread management techniques. The threads must be scheduled competently and synchronization between the threads is also essential. Equally important is the efficient use of various memories, viz. texture memory, constant memory, shared memory, global memory and local registers. All these can contribute in a significant performance improvement.

A challenging requirement is to parallelize the algorithms to be implemented in GPU. Significant efforts are needed to refactor the sequential algorithms to achieve the necessary computational margin in parallel scenario.

The effectual bandwidth of the computations should be considered as a metric when measuring performance and optimization benefits. To achieve improved performance gain, data transfer between the CPU and the GPU is to be minimized.

The GPU used for this design set up is nVIDIA Tesla C2075 with 448 CUDA cores [13].

## 6. DESIGN APPROACH

The test image which is discussed here for urban area detection is a panchromatic image of Cartosat-1 with 2.5 meter resolution and the corresponding image for water body detection is an AWiFS image of Resourcesat-2 with 12-bit radiometric resolution and data acquisition in four spectral bands, B2, B3, B4 and B5 (0.52-0.59; 0.62-0.68; 0.77-0.86; 1.55-1.70 μm) with 56m spatial resolution. The test images are of sizes, 4000x4000 pixels and 2000x2000 pixels.

The automated feature extraction algorithm had essentially four techniques to be implemented on the image so as to identify the region of interest. These techniques were implemented on a kernel mask of 5x5 and extended to the entire image. This modus operandi was extremely time-consuming and hence it was indispensable to realize alternate efficient parallelization strategy to accelerate the entire processing chain.

### 6.1. Design Approaches with CPU

The initial design technique was to employ conventional sequential code in a single core processor. Highly computation intensive steps discussed earlier will be exceedingly time consuming, as evident from the succeeding table (Table 1).

The execution speed for all the steps in the automated feature extraction algorithm can be increased by the exploitation of the inherent parallel resources of a multi-core processor. The current CPU is an 8-core processor and all its cores can be exploited effectively by dividing any application or algorithm into modules that are independent of each other and executing each portion in each core. This can be done using the concept of POSIX or pThreads.

In this algorithm , the steps of LoG filtering, zero crossing detection followed by feature extraction using standard deviation were duplicated on the raw image, once with the filter of high sigma and then with the filter of low sigma. Since these were independent of each other, these were executed in parallel using the multi-core processor. This involved invoking two parallel threads, each independently executing the steps. The results of both were combined to extract the features and edges resulting in a significant performance improvement as detailed below. The time gain obtained by the parallel approach employing CPUs is detailed in Table 1.

Table 1. Comparison of approaches CPU

| Image Details | Conventional CPU approach | Parallel approach employing CPUs | |
|---|---|---|---|
| | Time (ms) | Time (ms) | % speed-up |
| Cartosat 1 2000x2000 | 5950 | 3420 | 74% |
| Cartosat 1 4000x4000 | 22030 | 12010 | 83.5% |
| AWiFS 2000x2000 | 42838 | 24500 | 74% |

## 6.2 Design Approaches with GPU

The input image available in CPU (host) memory was copied to GPU (device). The entire processing is carried out in GPU and the detected region of interest is sent back to CPU, after the processing is completed.

The following sections discuss about the GPU implementation of the Automated Feature Extraction algorithm employing various optimization approaches and analysis of the throughput achieved in each case by taking different parameters into consideration.

### 6.2.1 Row-wise Parallel Approach

An approach that involved assigning a row of pixels to each thread was attempted. In this approach, an entire row of the image was processed by an individual thread. The total number of threads was chosen to be the total number of rows, with each thread handling each row of pixels. The maximum number of threads/block is 1024 and the optimum number is 512, which was chosen and the number of blocks exercised depended on the image size. The threads were organized into kernel launch parameters as given in Table 2. This approach has proved faster compared to the parallel CPU approach and has yielded a considerable throughput that is evident in Table 3.

Table 2. Kernel Launch Parameters for Row-wise Parallel Approach

| Image Size | No of threads | No of Blocks | No of Grids |
|---|---|---|---|
| 2000x2000 | 512 | 4 | 1 |
| 4000x4000 | 512 | 8 | 1 |

Table 3. Comparison of Row-wise Parallel GPU Approach with CPU

| Image Details | CPU Execution Time (ms) | GPU Execution Time (ms) | % Speed-up |
|---|---|---|---|
| Cartosat 1 2000x2000 | 3420 | 1000 | 242 |
| Cartosat 1 4000x4000 | 12010 | 1400 | 757.86 |
| AWiFS | 24500 | 1560 | 1470.51 |

| 2000x2000 | | | | |
|---|---|---|---|---|

### 6.2.2 Pixel-wise Parallel Approach

Efficient utilization of GPU resources can be achieved by designing the flow of execution in such a way that the logic will be implemented on all the pixels of the image at a time, in parallel. The total number of threads was chosen to be equal to the total number of pixels in the image (number of rows multiplied by number of columns in the input image), with each thread handling each pixel.

A new concept of global thread indexing has been employed in this approach which can be visualized below in Figure 3 (explained with reference to an input image of 2000x2000 pixels).

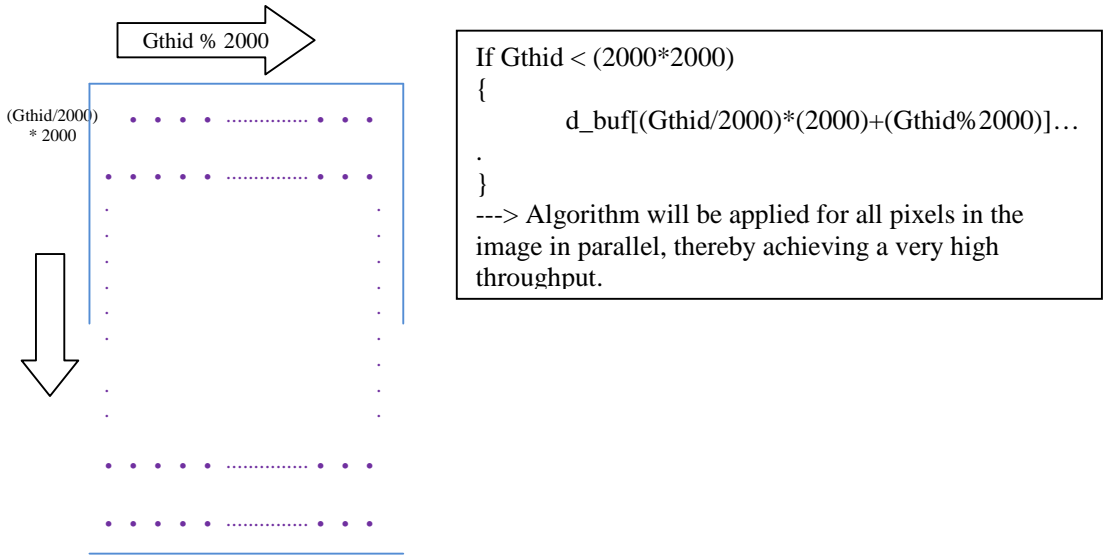

Figure 3. Global thread indexing

Using the equation $(Gthid/2000)*(2000) + (Gthid\%2000)$ for the image of size 2000x2000, all the pixels in the image were accessed and the logic was executed in parallel for all the pixels.

### 6.2.3 Selection of Kernel Launch Parameters

The number of threads and the number of blocks is chosen so as to minimize the execution time. Various threads and block combinations viz. 256/16384, 512/8192 and 1024/4096 were tested which yield different execution times as shown in the table (Table 4) below

Table 4. Kernel Launch Parameters for Pixel-wise Parallel Approach

| Image Size | No. of threads | No. of blocks | No. of grids | Execution Time (ms) |
|---|---|---|---|---|
| 2000x2000 | 256 | 16384 | 1 | 457 |
| | 512 | 8192 | 1 | 446 |
| | 1024 | 4096 | 1 | 520 |

The above table clearly depicts that the execution time was minimum for a block size of 512 threads and hence was finalized.

### 6.2.4 Optimization Strategies Incorporated

The entire operations involving the edge detection and detecting the regions of interest were done with a single kernel call from the host. The optimization strategies adopted are detailed in Table 5

Table 5. Optimization Strategies

| Optimization Strategies Adopted | Description |
|---|---|
| Parallelization Strategy | Involves assigning one pixel per thread, so that the number of threads is equivalent to the number of pixels in the input image |
| Threads per block | Number of threads per block was chosen as 512 |
| Multiple Kernels | For improved performance margins, multiple kernels were invoked within the device, to reduce the load on each kernel considerably |
| Block Synchronization | Synchronization between different row-wise and column-wise operations was carried out using syncthreads () |

## 7. PERFORMANCE EVALUATION

The developed GPGPU application was implemented in the following workstation:
    CPU    : HP Z-800 Intel Xeon 8 core E5620 @2.4Ghz
    GPU    : nVidia Tesla C2075 with 448 CUDA cores

Subsequent to the analysis of a variety of design approaches in CPU and GPU, we have zeroed upon the following design configuration which yielded the maximum performance margin.
Invoking multiple kernels with a pixel-wise parallel approach in GPU having 512 threads per block provided us the best approach. This section provides the results for various test images with reference to kernel execution timings, data transfer timings and the speed up attained.

The succeeding table (Table 6) provides the data transfer time (host to device and device to host transfer times combined together) and the kernel execution time for the test images

Table 6. Details of GPU execution timings

| Image Details | Data Transfer Time (ms) | Kernel Execution Time (ms) |
|---|---|---|
| Cartosat 1 2000x2000 | 10 | 69 |
| Cartosat 1 4000x4000 | 44 | 372 |
| AWiFS 4000x4000 | 6 | 825 |

This is the most efficient approach compared to its counterparts i.e., conventional CPU approach, parallel approach using CPU (pThreads) and the GPU approach of on row of pixels per thread. The efficiencies of various approaches are tabulated in Table 7. The speed up obtained is used as the measure of throughput.

Table 7. Comparison of Pixel-wise parallel GPU approach with pthreads in CPU

| Image Details | CPU --- pthreads Time(ms) | GPU --- Pixel-Wise Parallel Time(ms) | % Speed-up |
|---|---|---|---|
| Cartosat 1 (2000x2000) | 3420 | 446 | 666.82 |
| Cartosat 1 (4000x4000) | 12010 | 990 | 1113.13 |
| AWiFS (2000x2000) | 24500 | 1182 | 1972.76 |

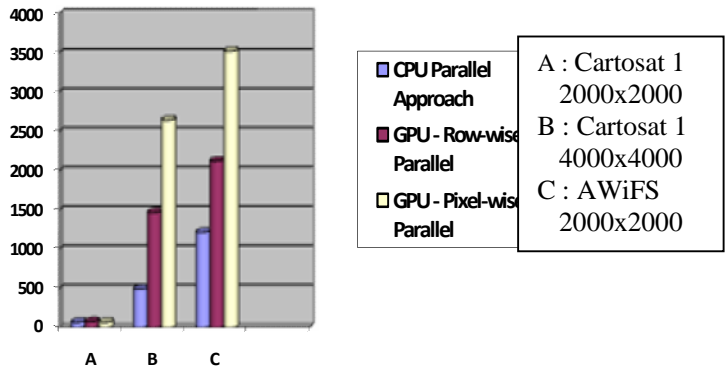

Figure 4. Representation of throughputs of various approaches

The following figures, Figure 5 - Figure 8, denote the Cartosat test image with 2000x2000 pixels for urban area detection, the features isolated in Figure 6, detected edges in Figure 7 and the final image showing the output of automated feature extraction which detected roads and buildings, in Figure 8. Figure 9 and Figure 10 denote the input and output of a segment of 4kx4k image. Figure 11 - Figure 14, depict the AWiFS image with water bodies, the extracted features and edges in Figure 12 and Figure 13 respectively and the final extracted water bodies in Figure 14.

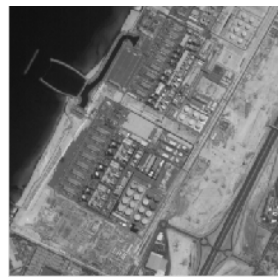
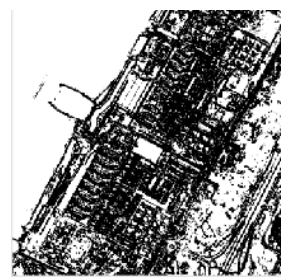

Figure 5. Input image for Urban Area Detection (2Kx2K)

Figure 6. Image with features isolated

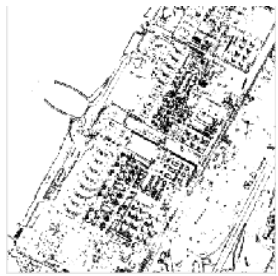
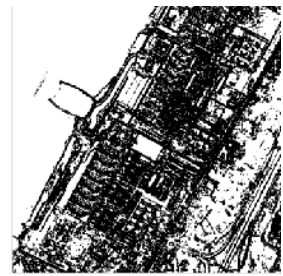

Figure 7. Image with extracted edgesFigure 8. Image with features extracted

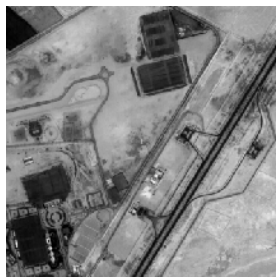
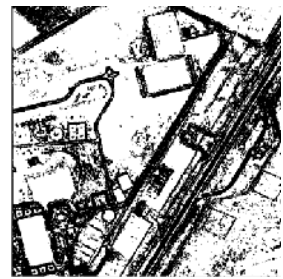

Figure 9. Input image for Urban Area Detection (4Kx4K)Figure 10. Image with features extracted

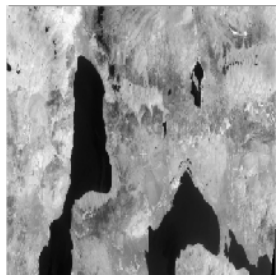
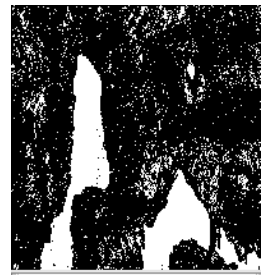

Figure 11. Input AWiFS image for water body Detection(2Kx2K)Figure 12. Image with features isolated

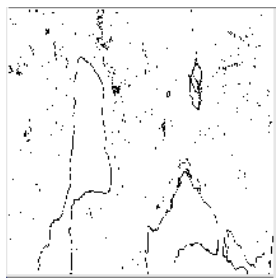
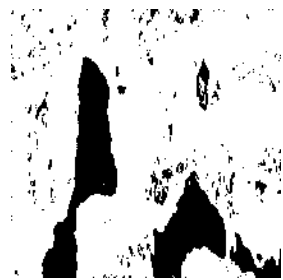

Figure 13. Image with extracted edgesFigure 14. Image with extracted water bodies

The efficiency of the parallelization strategy adopted can be measured by comparing the GPU execution time with the CPU execution time. A maximum percentage improvement of 1972% as compared to parallel approach using multi-core CPU was obtained with this parallelization methodology.

## 8. CONCLUSION

The Graphical Processing Units are exceedingly capable parallel processors which acquires data from the CPU to carry out repetitive calculations on huge volumes at an extremely rapid rate. The processed data would then be sent back to the host. The discussed GPU design configuration has provided a noteworthy performance improvement compared to even a multi-core CPU processing. This developed algorithm has successfully extracted the features and edges from panchromatic as well as multi-spectral images with relatively large resolutions as well. This can be further extended to extract any specific feature including roads, buildings, water bodies, snow, cloud etc., from a panchromatic image or a multispectral image using textures. This will help in isolating and classifying diverse features with enhanced accuracy.

## ACKNOWLEDGMENTS

The authors desire to acknowledge the support and excellent guidance made available by Dr. V. K. Dadhwal, Director, NRSC in successfully implementing this work. The support provided by Sri.K.Abdul Hakeem, Water Resources Division, NRSC and Ms.P.K.Saritha, SDAPSA, NRSC is also duly acknowledged.

**Authors**

Ms. K. Phani Tejaswi did her B.Tech in Electronics and Communications Engineering,from JNTU College of Engineering, Hyderabad. She is currently working as Scientist in Data Processing Area in National Remote Sensing Centre, Indian Space Research Organization (ISRO). She is involved in algorithm development and image processing in GPGPU.

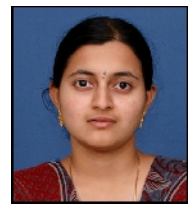

Mr. D. Shanmukha Rao did his M.Sc, Mathematics from Andhra University, Visakhapatnam. He is currently working as a Scientist in Data Processing Area in National Remote Sensing Centre, Indian Space Research Organization (ISRO). He is involved in algorithm development for data processing and image analysis.

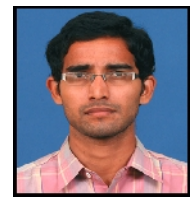

Ms. Thara Nair did her M. Tech, Control Systems Engineering from College of Engineering, Trivandrum. She worked as Scientist in Vikram Sarabhai Space Centre, Indian Space Research Organization from 1997 - 2010. She is currently working as Scientist in Data Processing Area in National Remote Sensing Centre, Indian Space Research Organization (ISRO). She is involved in High Performance Computing and algorithm development for image processing applications using GPGPU.

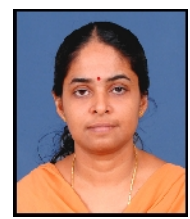

Mr. A.V.V. Prasad did his M.Sc, Physics (Electronics) from Andhra University, Visakhapatnam in 1985. He is presently the Group Head of Microwave Remote Sensing and Global Data Processing Group of National Remote Sensing Centre, Indian Space Research Organization (ISRO). He was involved in the installation of Remote Sensing satellite data reception systems and test systems like Advanced Front End Hardware units (AFEH), Serializer systems, Data logging systems etc., for different satellites like IRS-1C, IRS-P4,IRS-P5, IRS-P6 , TES etc., and a data capturing facility at Arctic Station, Svalbard and Antarctica.

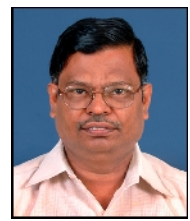